# Dimerization and Fusion of Two $C_{60}$ Molecules

Narinder Kaur[1], K. Dharamvir[2] and V. K. Jindal[2*]

[1]Department of Applied Sciences, Chandigarh College of Engineering and Technology, Chandigarh -160026, India

[2]Department of Physics, Panjab University Chandigarh -160014, India

**Abstract**

We investigate the dimerization and fusion of $C_{60}$ molecules to form various $C_{60}$ dimers when pushed against each other at several inter-molecular distances. We study the stability of this dimerized $C_{60}$ molecule based on its binding strength provided by intramolecular interactions. Tersoff potential, which is considered to represent intramolecular interactions well, has been used to calculate potential energy at these distances of separation and for all possible orientations of the molecules. We observe that several minimum energy configurations exist at various distances between the $C_{60}$ molecules. Our calculation shows that apart from the dumbbell structures, many interesting composite phases also result, such as fused, peanut and carbon nanotubes of geometry (5,5) of length 11.84 Å and (10,0) of length 12.30Å.

## 1 Introduction

The icosahedral 60-atom carbon clusters popularly known as bucky balls, have been subjected to much investigation for the past two decades. They form molecular crystals with weak intermolecular bonding, adequately represented by Van der Waals interactions [1-2]. In this crystalline state, at ambient temperature, the bucky balls are free to change their mutual orientation or even to rotate around the molecular centers while preserving a perfect crystalline lattice order. At temperature below 260K however, the orientational freedom gets frozen. A gentle push, hydrostatic pressure, excitation by light or other factors can promote a stronger covalent bonding between the $C_{60}$ molecules thus allowing them to share some of their electrons [3-4]. This process leads to formation of dimers, polymer like chains (pearl necklaces) or rigid two- and three -dimensional networks and may dramatically change the electronic and optical properties of the bulk material. Numerous fascinating properties of a collective nature are also expected to be displayed by these formations.

---

* Author with whom correspondence be made. Email: jindal@pu.ac.in

It has been known for some time now [4] that on applying high pressure and temperature new phases appear in which equilibrium distance between nearest neighbor $C_{60}$ molecules get shortened from 9.9 Å (in the crystalline state of individual bucky ball molecules) to about 9.0 Å. One obtains a dimer phase when $C_{60}$ solid is cooled rapidly from 450K to 77K [5]. Similarly, chain and layered polymer phases have been produced by cooling the $C_{60}$ solid slowly from 450K to 77K. The phase purity has been checked by X-Ray diffraction. The dimer phase is metastable and changes gradually to chain polymer phase.

Dimerization and polymerization of $C_{60}$ has also been observed under ion irradiated thin film samples of $C_{60}$ on Si and quartz substrates studied at controlled fluences, when exposed to swift heavy ions of hundreds of MeV[6]. Iijima and his coworkers [7-8] while heating nano-peapods (rows of bucky balls inside single walled carbon nanotubes) observed that inside the nanotube, the bucky balls started to coalesce at $800^0$C and, finally, completely transformed to a single-wall nanotube at $1200^0$C.

The bond between two adjacent bucky ball monomers in dimers or polymers is either $sp^3$ like bond (single bond [9] or a 2+2 cycloaddition bond) [10-11] or $sp^2$ like bond [8]. These types of bonds have been experimentally identified. X-ray powder diffraction studies established [12] that in the dimer molecule formed through a single C-C interfullerene bond the center to center distance d, between two bucky balls is 9.1Å while that in a cyclo-added dimer is 9.3Å. $sp^2$ like bonding is seen during the coalescence of two bucky balls giving rise to very strong intercage bonding with d ~ 8.53 Å. First principles molecular dynamical relaxation has been performed to study seven different configurations of a $C_{60}$ dimer by Adams et. al. [10]. They showed the cycloadded dimer to be the most stable dimer phase whereas other considered dimer structures were shown as unbound structures. Menon et.al.[11] have used generalized tight-binding molecular dynamics technique to study various $C_{60}$ dimer geometries and have suggested the cycloadded dimer structure to be the minimum energy dimer structure.

Porezag et.al. [13] have used density function based nonorthogonal tight binding method to study the structure, energetics and vibrational properties of five different $(C_{60})_n$ oligomers (n=2,3 and 4). For various values of b, the $C_{120}$ molecule was minimized by allowing 118 atoms to move and keeping the two atoms making dimer bond, fixed. Cycloaddition bond has been suggested as the linkage style in a $C_{60}$ dimer molecule with binding energy of 1.20eV.

Peapods have also been studied by way of molecular dynamics simulations and it has been shown that 22 successive Stone-Wales (SW) transformations (90° bond rotation within the plane of a $sp^2$-carbon network) are a topologically acceptable pathway for the transformation of a cycloadded dimer molecule to a perfect $C_{120}$ buckytube [14-17]. However, the ordinary molecular dynamics study for the process was not applicable. For the whole coalescence process, the energy barrier has been estimated to be 8-12eV [8]. Marcos et.al. [18] find in their calculations that a thermal road to the annealing process requires much less expenditure of energy and have proposed 24 stable $C_{120}$ isomers in SW pathway including the initial [2+2] and final state $C_{120}$ nanotube.

Another approach is to study the collisions between two $C_{60}$ molecules [19]. By using molecular dynamics simulations it has been shown that dumbbell-shaped $(C_{60})_2$ dimer with almost intact cages could be formed at low collision energy (~21.5eV) and coalesced dimer with symmetrically distorted cages at higher incident energy (~52eV) and when the collision energy was high enough (~400eV) to overcome the fusion barrier, the two colliding $C_{60}$ molecules fused to form one large cage-cluster, the $C_{120}$ molecule.

Solids formed out of dimerized $C_{60}$ molecules have also been studied theoretically using model potential approach [20-21]. These solids can have orthorhombic, tetragonal or monoclinic structures. The pressure and temperature dependent properties such as bulk modulus, lattice and orientational structure, phonon dispersion relations, Gruneisen parameters, heat capacity and entropy have been studied by us, considering the rigid dimer molecule to be singly bonded [20]. The possible crystal packings of dimerized $C_{60}$ molecules were studied by minimizing the lattice energy with a bond charge intermolecular potential model, when the two $C_{60}$ monomers are cycloadded [21]. The study on the $C_{120}$ cluster produced by Laser desorption of fullerene films, to measure ion mobility proposed several $C_{120}$ fullerene cages with different gross shapes [22].

On the basis of theoretical and experimental work done up till now it emerges that several results have been reported in literature which are based on different techniques. A comprehensive and systematic study of the dimerization process of two bucky balls has not been reported by a single approach. The detailed study on the requisite conditions and the types of bonds formed between two $C_{60}$ molecules when brought closer to each other is desirable.

In this paper we study the various possible structures of $C_{120}$ (henceforth a "bonded double $C_{60}$" will be called dimer molecule) obtained by bringing together two $C_{60}$ cages. The structures obtained have been analyzed to obtain the number of intercage bonds, bond energies, lengths and several other characteristics. We have used a theoretical model in which the interaction between bonded carbon atoms is governed by the Tersoff potential [23]. In section 2, the form of this potential is given, which has been used to calculate the binding energy of two bucky balls at various intercage distances for all possible orientations. In section 3, the numerical results for the dimer structures obtained by using our theoretical model are presented. The structures are discussed individually in section 4. Conclusions have been drawn about several stable dimer structures in the last section of the paper.

## 2 Theoretical procedure

The form of the Tersoff potential for calculating potential energy between $i^{th}$ and $j^{th}$ bonded carbon atoms at $r_{ij}$ distance apart is described below in section 2a. In section 2b we describe the procedure for obtaining different structures of the dimer molecule.

### a) *The model Potential*

The Tersoff potential has been successfully used for the modeling of intramolecular chemical bonding in a wide range of hydrocarbon molecules [24], diamond and graphite lattices as well as carbon nanotubes [25]. The results obtained for elastic constants and phonon dispersion, were in good agreement with experiments and with ab initio calculations (for defect energies). The potential is able to distinguish among different carbon environments, fourfold $sp^3$ bond as well as threefold $sp^2$ bond.

The potential consists of a pair of Morse-type exponential functions with a cut-off function $f_c(r)$.

$$V_{ij} = f_c(r_{ij})[a_{ij}V_R(r_{ij}) + b_{ij}V_A(r_{ij})] , \qquad (1)$$

Where $V_R(r)$ and $V_A(r)$ are repulsive and attractive force terms, respectively.

$$V_R(r_{ij}) = Ae^{-\lambda_1 r_{ij}} , \qquad (2)$$

$$V_A(r_{ij}) = -Be^{-\lambda_2 r_{ij}} . \qquad (3)$$

$f_c(r)$ is a function used to smooth the cutoff distance. It varies from 1 to 0 in sine form between R-D and R+D, D being a short distance around the range R of the potential

$$f_c(r) = \begin{cases} 1, & r < R - D \\ \frac{1}{2} - \frac{1}{2}\sin\left[\frac{\pi}{2}(r-R)/D\right], & R - D < r < R + D \\ 0, & r > R + D \end{cases} \quad (4)$$

The other functions in equation 1 are,

$$b_{ij} = \frac{1}{\left(1 + \beta^n \xi_{ij}^n\right)^{\frac{1}{2n}}}, \quad (5)$$

where, $\xi_{ij} = \sum_{k \neq i,j} f_c(r_{ik}) g(\theta_{ijk}) e^{[\lambda_3^3 (r_{ij} - r_{ik})^3]}$ \quad (6)

Here $\theta_{ijk}$ is the bond angle between ij and ik bonds as shown in Fig 1. The state of the bonding is expressed through the term $b_{ij}$ as the function of angle between bond *i-j* and each neighboring bond *i-k* and

$$g(\theta) = 1 + \frac{c^2}{d^2} - \frac{c^2}{[d^2 + (h - \cos\theta)^2]} \quad (7)$$

Further,

$$a_{ij} = (1 + \alpha^n \eta_{ij})^{\left(-\frac{1}{2n}\right)} \approx 1 \quad (8)$$

$$\eta_{ij} = \sum f_c(r_{ik}) e^{\left(\lambda_3^3 (r_{ij} - r_{ik})^3\right)}, \text{ when } \alpha \text{ and } \lambda_3 \text{ are taken as 0} \quad (9)$$

$a_{ij} \neq 1$ only if $\eta_{ij}$ is exponentially large, which will occur for atoms outside the first neighbor shell. Initially coordination number of each atom (say $i^{th}$) is 3 ($k_1$, $k_2$, $k_3$) in a $C_{60}$ molecule, but when two neighboring bucky balls come close to each other for intermolecular bonding this number increases depending upon the intermolecular distance. We observe that the parameter $\lambda_3$ taken 0 for carbon systems plays a major role in controlling the attractive forces between the contacting carbon atoms, when coordination number increases from 3 to 5 and higher during dimerization reaction. We have presented our results using the value of this parameter as zero.

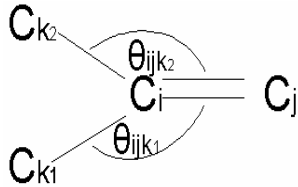

**Fig 1:** Showing a set of four neighboring carbon atoms

Using this potential, composite energy of all the atoms of the system is given by $E_b$ which is written as

$$E_b = \frac{1}{2}\sum_{ij} V_{ij} \tag{10}$$

The sum in Eq.10 includes all the atoms in each of the molecules. All the parameters appearing in the expressions for potential have been tabulated in Table I.

**Table I:** Showing original and modified parameters of the potential.

| Tersoff Parameters | Original [26] | Modified |
|---|---|---|
| A(eV) | 1393.6 | 1380.0 |
| B(eV) | 346.7 | 349.491 |
| $\lambda_1$(Å$^{-1}$) | 3.4879 | 3.5679 |
| $\lambda_2$(Å$^{-1}$) | 2.2119 | 2.2564 |
| $\beta$ | 1.57 x 10$^{-7}$ | 1.57 x 10$^{-7}$ |
| n | 0.72751 | 0.72751 |
| c | 38049. | 38049.0 |
| d | 4.3484 | 4.3484 |
| h | -0.57058 | -0.57058 |
| R(Å) | 1.95 | 1.95 |
| D(Å) | 0.15 | 0.15 |

In our case we deal with a composite molecule $C_{120}$ consisting of two $C_{60}$ balls. The summation in eq.10 therefore runs from 1 to 120 atomic indices.

For the intermolecular interactions between the non bonded i$^{th}$ and j$^{th}$ carbon atoms, Van der Waals interaction potential is used which is given by the expression

$$V_{ij} = -\frac{A}{r_{ij}^6} + B \exp(-\alpha r_{ij}) \tag{11}$$

Where A, B and α are the interaction parameters. These parameters have been tabulated for carbon-carbon interactions in the literature [20]. However in comparison to the Tersoff potential, at the distances of consideration, the intermolecular interaction potential generated by Eq.(11) is numerically insignificant so it has not been included in the present calculation.

**b)** *Adjustment of Potential Parameters*

Two types of bond lengths determine the coordinates of 60 carbon atoms in $C_{60}$ molecule. Single bond $b_1$, joining a hexagon and a pentagon is of length 1.45Å and double bond $b_2$ joining two hexagons is shorter, having length 1.40Å [27]. By using the parameters given by Tersoff, the structure was allowed to minimize using the potential model as given in the earlier section. In this way, $b_1$, $b_2$ and bond angles were varied to obtain minimum energy configuration. By doing this, $b_1$ and $b_2$ were obtained to be 1.46Å and 1.42Å with binding energy 6.72eV/atom as given in Table III.

In order to reproduce the bond lengths and the binding energy of a $C_{60}$ molecule in closer agreement with the experimental results of Dresselhaus et.al.[27], the potential parameters given by Tersoff [26] had to be modified. It is found that the first four Tersoff parameters A, B, $\lambda_1$, $\lambda_2$ are the effective parameters to get appropriate binding energy and bond lengths so only these were modified. In Table I we gave the modified parameters as well as the original potential Parameters. The new bond lengths and energies have been given in Table III. The modified parameters have been used by us to obtain minimum energy configurations.

**Table III**: Comparison between the calculated and experimental binding energy and bond lengths of a $C_{60}$ molecule with original and modified parameters.

|  | Calculated | | Experimental [27] |
|---|---|---|---|
|  | With Tersoff Parameter | Present work |  |
| Binding energy (eV/atom) | -6.73 | -7.17 | -7.04 |
| Bond lengths (Å) | 1.46,1.42 | 1.45,1.41 | 1.45,1.40 |

**c) Application to Dimer Structure**

We adopt a procedure (relaxation) where the initial configuration consists of two bucky balls with a certain mutual orientation at a certain distance apart such that the bucky balls are within chemical bonding range. The atomic coordinates of the 120 atoms are then adjusted one by one to obtain a configuration with lower energy. The cycle is repeated many times till minimum energy is obtained. The initial configuration was fixed by the following criteria. We identify 16 different configurations of two $C_{60}$ molecules on the basis of orientational differences, which lead to distinct dimer structures. These are shown in Table II. Various combinations of single bond SB, double bond DB, corner atom, hexagonal face HF and pentagonal face PF with each other, make different starting configurations of the two facing molecules, such as single bond of one molecule facing single bond (SB-SB) or single bond of one molecule facing double bond of the other molecule (SB-DB) and so on.

**Table II:** Showing sixteen possible orientations of two facing $C_{60}$ molecules

| S.No | Orientation Symbol | S.No | Orientation Symbol |
|---|---|---|---|
| 1 | SB-SB | 9 | DB-HF |
| 2 | SB-DB | 10 | C.Atom - C.Atom |
| 3 | SB-C. Atom | 11 | C.Atom-PF |
| 4 | SB-PF | 12 | C.Atom-HF |
| 5 | SB-HF | 13 | PF-PF anti parallel |
| 6 | DB-DB | 14 | PF-PF parallel |
| 7 | DB-C.Atom | 15 | PF-HF |
| 8 | DB-PF | 16 | HF-HF |

The binding energy of two bucky balls w.r.t. intercage distance d has been shown in Fig 2 for two orientations i.e. DB-DB, in which double bonds of the two balls face each other and PF-PF, in which pentagonal faces of the two balls face each other anti-parallel. Incase of DB-DB orientation dimer phase starts from 9.0Å at which we obtained a minimum energy dimer structure with single covalent bond as the intercage bond as shown in fig 3. Another dimer structure is obtained at 8.6Å with cycloaddition bond as the intercage bond. We have been able to obtain one more stable new dimer structure at 7.9Å with five covalent bonds as the intercage bonds giving rise to a fused structure. However for the PF-PF orientation we obtained single minimum at 8.3Å which give rise to structure 7 as shown in Fig 3.

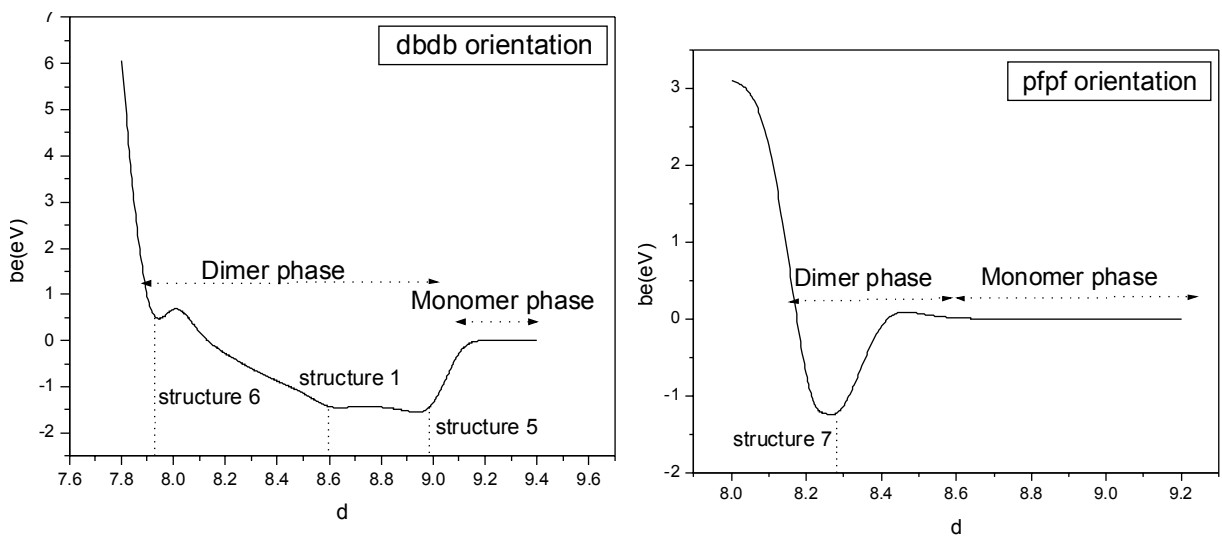

**Fig 2** Binding energy of two bucky balls with DB-DB and PF-PF orientation at various intercage distance

Similar plots for other possible orientations were studied and it was found that initial intercage distances are usually from 7.9Å to 9.1Å and yield the dimer structures. Several other minimum energy configurations are possible to obtain using distorted $C_{60}$ molecules (i.e. by opening some of the on-cage bonds). Within our treatment, this is the only way to arrive at the final states of carbon nanotube $C_{120}$ and the peanut. The cage-opening represents thermal activation as has also been described by Marcos et. al.[18].

## 3. Numerical Results

The dimers obtained were categorized depending upon their bonding schemes and are shown in Fig 3. we find that the possible three classes are a) structures with very few bonds, not significantly disturbing carbon atoms other than those involved in intercage bonding; b) Fused structures – those in which contact atoms have some of the original $C_{60}$ bonds broken and new bonds formed- mixture of $sp^2$-$sp^3$; c) Coalesced structures- those with all of the carbon atoms finally attaining $sp^2$. Further features of these structures are discussed in the next section. The numerical results are summarized in Table-IV.

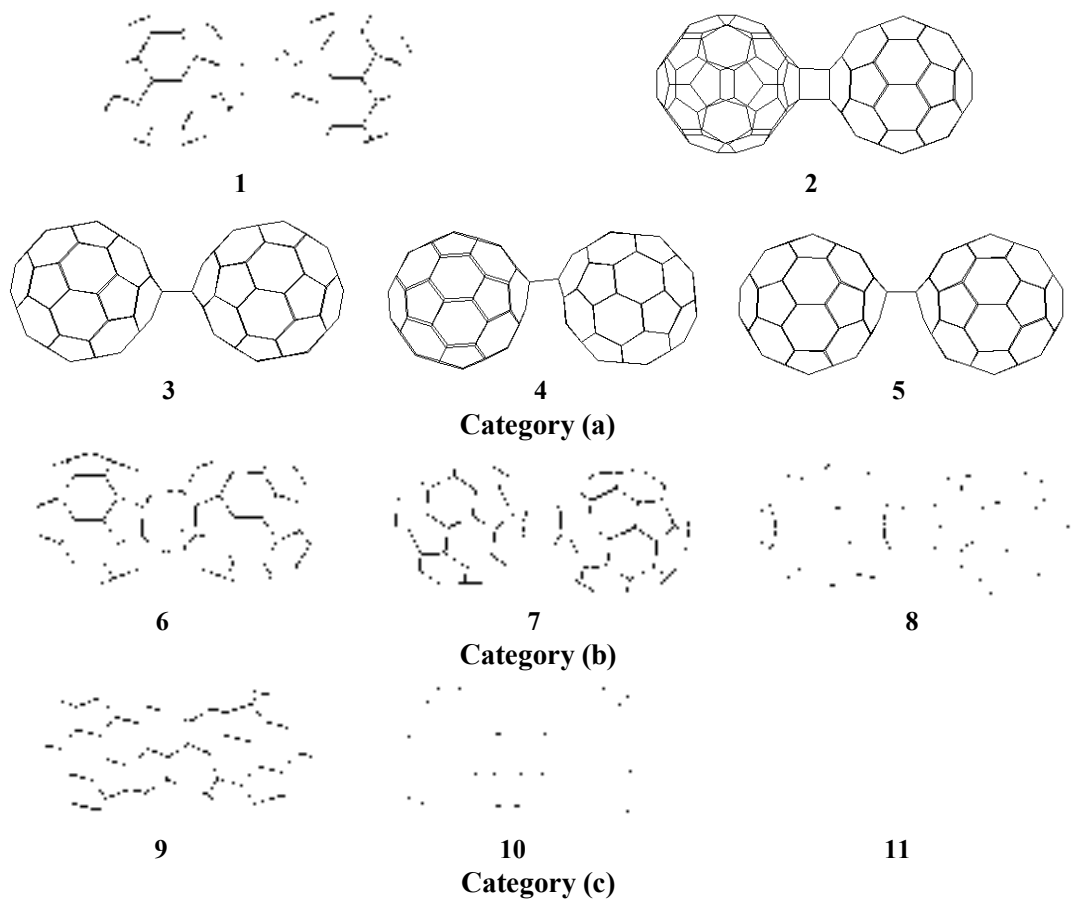

1  2

3  4  5

**Category (a)**

6  7  8

**Category (b)**

9  10  11

**Category (c)**

**Fig. 3** Various structures obtained after relaxation under Tersoff potential. Set of diagrams in category a) shows $sp^3$ like intercage bonding, b) shows a mixture of $sp^3$ and $sp^2$ like intercage bonding and c) shows pure $sp^2$ like intercage bonding.

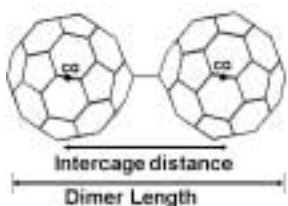

**Fig 4**: showing intercage distance and dimer length

**Table-IV** The orientation and minimized energies of the eleven $C_{60}$ dimer structures

| St. No. | Structure as shown in Fig 3 | Orientation | Initial Center to Center Dis. (Å) | Initial energy (eV) | Minimized energy (eV) | No. of inter cage bonds | Intercage Bond length (Å) | Final Center to Center Dis. (Å) | Dimer Length (Å) |
|---|---|---|---|---|---|---|---|---|---|
| 1 | Dumbbell | DB facing DB | 8.6 | .78 | -1.51 | 2 | 1.54(2) | 8.903 | 15.95 |
| 2 | Cycloadded | SB facing DB | 8.0 | 44.59 | -1.86 | 2 | 1.53, 1.54 | 8.82 | 15.83 |
| 3 | Single bonded-1 | C-atoms Facing | 8.4 | 3.41 | -2.165 | 1 | 1.48 | 9.08 | 16.196 |
| 4 | Single bonded-2 | DB facing PF | 8.5 | 7.44 | -1.07 | 1 | 1.486 | 8.68 | 15.68 |
| 5 | Single bonded-3 | DB facing DB | 9.0 | 0.9 | -1.71 | 1 | 1.485 | 8.975 | 16.016 |
| 6 | Fused – 1 | DB facing DB | 7.9 | 88.63 | -.14 | 5 | 1.49(2) 1.51 1.33 1.56 | 7.87 | 14.92 |
| 7 | Fused – 2 | PF facing PF anti-parallel | 8.3 | 22.63 | -2.22 | 2 | 1.523(2) | 8.381 | 15.27 |
| 8 | Fused – 3 | SB facing PF | 8.5 | 5.47 | -0.52 | 2 | 1.567(2) | 8.75 | 15.71 |
| 9 | Peanut | Open Hf facing Closed Hf | 8.1 | 10.61 | -16.3 | 6 | 1.39 each | 8.50 | 15.44 |
| 10 | Armchair nano tube | Open PF facing open PF | 8.0 | 130.08 | -33.64 | 6 | 1.41 each | 3.92 | 11.84 |
| 11 | Zigzag nanotube | *$C_{120}$ isomer | - | 26.93* | -35.10 | 6 | 1.41 each | 6.29 | 12.30 |

* Facing pentagons of two $C_{60}$ isomers are opened and these monomers are brought closer. We have minimized the $C_{120}$ isomer with our potential model, so initial energy required is much more than quoted here.

In Table IV we defined the center to center distance, referred to as the intercage distance, as the distance between the center of gravity of the first 60 atoms and that of the remaining 60 atoms, originally belonging to the two bucky balls as shown in Fig 4. The dimer length is defined as the end to end axial distance between the two balls. The minimum energy represents the energy of the dimer molecule by allowing all the 120 atoms to relax. In reality only a few atoms closer to the two molecules relax appreciably as shown in Fig 3. The structures obtained after relaxation from open cages require, in addition to proximity, some extra energy, which could be provided by temperature.

We have not considered the energetically stable cages i.e. Cage-$T_d$, Cage-$C_{1H}$, Cage-$C_2$ studied by Esfarjani et.al [28] or toroidal cage form of point group $D_{5d}$ of $C_{120}$ structures studied by Ihara et.al. [29], because our procedure of obtaining a $C_{120}$ structure was by compressing two $C_{60}$ monomers for various possible orientations so that the two balls retain their individuality at least by 50%, after the dimer formation. For the above mentioned structures the two balls completely lose their identities, so these structures were not studied.

**4. Discussion of the obtained dimer structures:**

In this section, we discuss the results i.e. Fig.3 and Table IV.

a) **Single bonded or cycloadded dimers**

The dimer structures under this category are formed when the initial intercage distances are between 8 Å to 9 Å. For different orientations, different bonding schemes result, as shown in Fig 3a. In **structure 1** the two balls are placed in a manner that double bonds of each of the two balls face each other. A ring type intercage bonding is there called 2+2 cycloaddition. Intramolecular double bond of each ball breaks and form parallel covalent intermolecular bonds of bond length 1.54 Å each, whereas the intramolecular bonds of this ring are 1.47 Å each. This type of bonding is the most talked about bonding in the $C_{60}$ dimers and polymers. The central $C_4$ unit that connects the neighboring bucky balls can be viewed as a cyclobutane fragment where every carbon atom is connected to four others. **Structure 2** is obtained when single bond of one ball faces the double bond of the other ball placed at around 8 Å distance apart. It is more stable than structure 1(see table IV) although their distortion at the interconnecting sites is similar. **Structure 3** is obtained when the carbon atoms of the balls face each other at around 8.4 Å. This type of bonding has been observed in alkali doped $C_{60}$ solids. In **Structure 4** double bond of one ball face pentagonal face of the other ball and in **structure 5** again double bonds of the two balls face each other but at 9 Å to form single covalent bond as the intercage bond. Structures 3 and 4 look similar but have different energies and initial orientations. In fact 3, with lower energy has shorter route during relaxation. Given enough time to relax, structure 4 also relaxes to 3. In this category structure 5 needs special mention as the intercage distance is larger than that in the other structures but it relaxes to be a very stable dimer (see Table IV).

b) **Fused dimers**

The dimer structures under this category are formed when the initial intercage distances are between 7.9 Å to 8.5 Å with $sp^2$ like intermolecular bonding (see Fig 3b). **Structure 6** is obtained from DB-DB orientation with 5 intercage bonds. **Structure 7** has been obtained when the pentagonal faces of the two $C_{60}$ molecules face each other anti-parallely. Two intramolecular bonds from the balls break and form two covalent intermolecular bonds. **Structure 8** is obtained when the single bond of one ball is facing pentagonal face of the second ball. An intramolecular bond break open the pentagon of one ball and the balls make two intermolecular bonds giving rise to a fused structure.

**c)   Coalesced dimers**

The dimer structures under this category were formed when the initial intercage distances were less than 8.2 Å. Different starting orientations; result in buckytube or peanut formation as shown in Fig 3c. In the resulting three structures all C-atoms are $sp^2$ bonded. We discuss these three coalesced structures individually.

**Structure 9** also referred as P66 [28] have been obtained by bringing two $C_{60}$ monomers in such a way that the hexagons of both the molecules are facing each other in which one has three loose bonds (the erstwhile single bonds) as shown in Fig 5a. Pure $sp^2$ type bonding can be seen at the inter-connecting sites. The balls fuse into each other making intermolecular $sp^2$ bonds and give rise to a very stable structure. The coalesced structure has been confirmed by laser desorption mass spectroscopy and its structure was assigned to a peanut shaped structure by comparison of the IR absorption spectrum, with theoretical spectra of five $C_{120}$ isomers by Strout et al [30]. Two more peanut structures have been studied by Esfarjani et.al. [28] i.e. P55 and P56. These structures have been observed to form in the Electron Beam-irradiated $C_{60}$ thin film as well as in the photo irradiated $K_xC_{60}$ film [31-33].

Other groups also found the coalesced dimers in aggregation following laser ablation of fullerene films, in collision between fullerene ions and thin films of fullerenes and in fullerene- fullerene collisions [34]. The Bucky balls have also been observed to coalesce during peapod formation reactions.

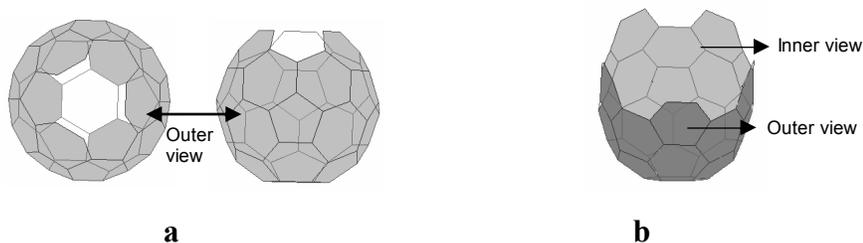

**a**  
**Fig 5a:** Front and Side view of opened cage             **b** for peanut formation

**5b**: cage opening for buckytube formation

**Structure 10,** the $C_{120}$ molecule in the form of Armchair Buckytube has been obtained by bringing together two $C_{60}$ monomers approaching each other in such a way that partially open pentagons are facing each other. This 'partial opening' has been shown in fig 5b. The contacting pentagon has all its single bonds cut (all five c-atoms allowed to move away from each other) so that each of these has only one remaining bond. As the two cages are now open they can fit into each other if brought very close, making new double bonds with other cage as shown in Fig 6. Pure $sp^2$ type bonding can be seen at the inter-connecting sites resulting in a nanocapsule of length 11.84Å. Theoretically breaking of the ten bonds of the two $C_{60}$ molecules seems easier way to open the cages, but it is believed that the isomerization mechanism is preferred as there is less expenditure of energy for the SW transformations, which can be viewed as bond rotations and require less energy than bond breaking. A sequence of only five SW-type bond rotations transforms a perfect $C_{60}$ molecule to a capped segment of a (5, 5) nanotube [18].

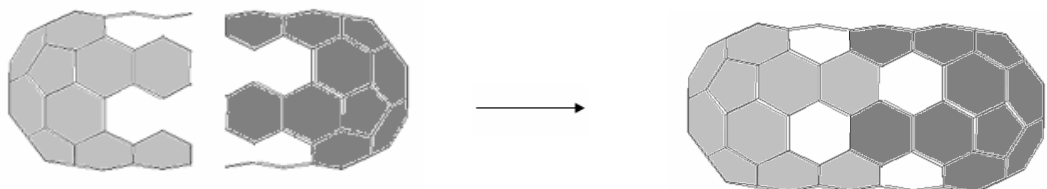

**Fig 6**: showing the seamless joining of two open $C_{60}$ molecules to form a buckytube

**Structure 11** *Zigzag Buckytube* was not attainable by any kind of cage opening as a precursor. Instead, we fed the assumed structure to the program and let it minimize. The resulting zigzag tube has a length of 12.30 Å, and is the most stable of the dimerized $C_{60}$ molecules found by us. Notice that the cap of zigzag nanotube is not the same as half of $C_{60}$ molecule. The terminating pentagon of this tube is surrounded by six hexagons as in the most stable $C_{60}$. However these hexagons are surrounded by PHHPHH sequence (P-pentagon, H-hexagon), instead of PHPHPH sequence as in the most stable $C_{60}$. This results in a cap appropriate for zigzag nanotube as shown in Fig 3c. All the ten relaxed structures as observed by us at various intercage distances have been shown in Fig.7. Structure 11 is not included as it has been obtained by a different method.

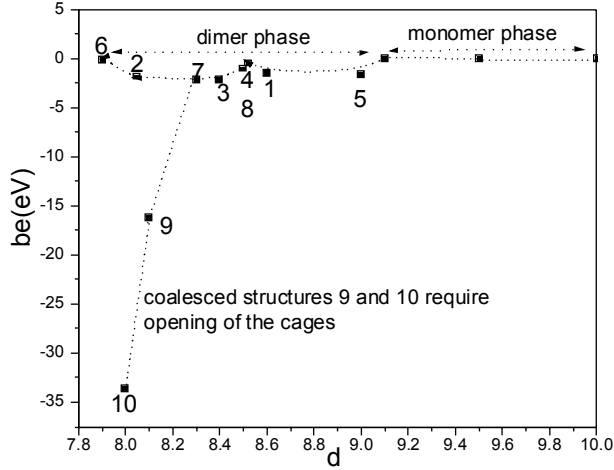

Fig 7: Energetically favorable structures at various intercage distances.

## 6 Summary and Conclusion

We have studied various forms of dimer $C_{60}$ obtained after squeezing together two bucky balls. In our calculations we see that, if $C_{60}$ molecules are considered rigid, then dimer structures 1, 3-5, 8 and 9 result in energies close to minima. The minima in energies are obtained by allowing the atoms of the molecules to relax, thus including the non-rigidity in molecules (see column 5 in Table IV). When non-rigidity of $C_{60}$ molecules is allowed then these structures are the energetically favorable structures (see column 6 in Table IV). Further, structures 2, 6 and 7 do not show any tendency to stabilize in rigid molecule model, they do stabilize when non-rigidity is considered (column 5 and 6 of Table IV). The binding energy of these structures comes out to be roughly of the order of 2eV which is in close agreement with the experimental estimate of 1.25eV [35]. However if opening of the cages in addition to non-rigidity is allowed then energetically most favorable structures are the coalesced structures (structures 9-11) with binding energy of roughly of the order of 34eV for the bucky tubes and 17eV for the peanut structure.

Structure 2 and 5 were obtained by Kim et.al. [36] as dianionic $(C_{60})^{-2}$ dimer phase while investigating sudden change in magnetic property of this phase in a rapidly cooled $AC_{60}$ samples (A-alkali metal). Fullerene coalescence, experimentally found in fullerene embedded single walled carbon nanotubes under heat treatment, has been simulated by minimizing the classical action for many atom systems for structures 10 and 11 by Kim et.al [8]. The initial state for the process has been taken as two $C_{60}$ molecules separated by 1nm. The synthesized inner tubes had their diameters ranging from 0.6-0.9 nm. Esfarjani et. al. [28] performed total energy minimizations for structures 1, 3, 4, 9 and 10 of Fig 3. Structure 7 and 8 were also studied by Choi.et.al [37] theoretically, using non local density functional theory while doing geometry optimizations on $C_{120}$. Structure 6 has never been reported so far.

We have minimized the above discussed structures after orientational positioning and partial opening of cages (as required for some dimer structures) have been done. So the calculated binding energies are lower than those quoted in the references 8 and 28. In reference 28 the most likely candidate for the experimentally found dimer structure i.e. Dumbbell structure has been reported as having positive binding energy, so their quoted energies are not in accordance with the experimental data available.

The bond energy for $sp^3$ like intercage bonding such as in structure 1 having intercage bond length 1.54Å and $sp^2$ like bonding as in structure 9 having intercage bond length 1.39Å have been estimated, which are of the order of 3.15eV/bond and 5eV/bond respectively. As bond strength is directly proportional to the bond energy of that bond so we conclude that the strength of $sp^2$ like bond is more than that of $sp^3$ like bond.

Some of the dimer structures have not been experimentally identified. The reason could be that the crystalline order restricts desired fusion or coalescence between two bucky balls when a $C_{60}$ solid is compressed. These structures can be formed by compressing $C_{60}$ molecules in gas phase.

It is interesting to observe all the known structures of dimerized and fused $C_{60}$ by following a uniform theoretical procedure adopted here. In this way, the possibility of losing any observable structure is very unlikely. Indeed, we find an unreported structure (structure 6) also. The results and inferences of this work provide motivation for experimentation on the $C_{60}$ dimer molecule forming the dimer solids. For the discussed dimer structures we propose to investigate the consequent structures of the dimer solids, in line with our earlier work [20].

**Figure 3**
Click here to download high resolution image

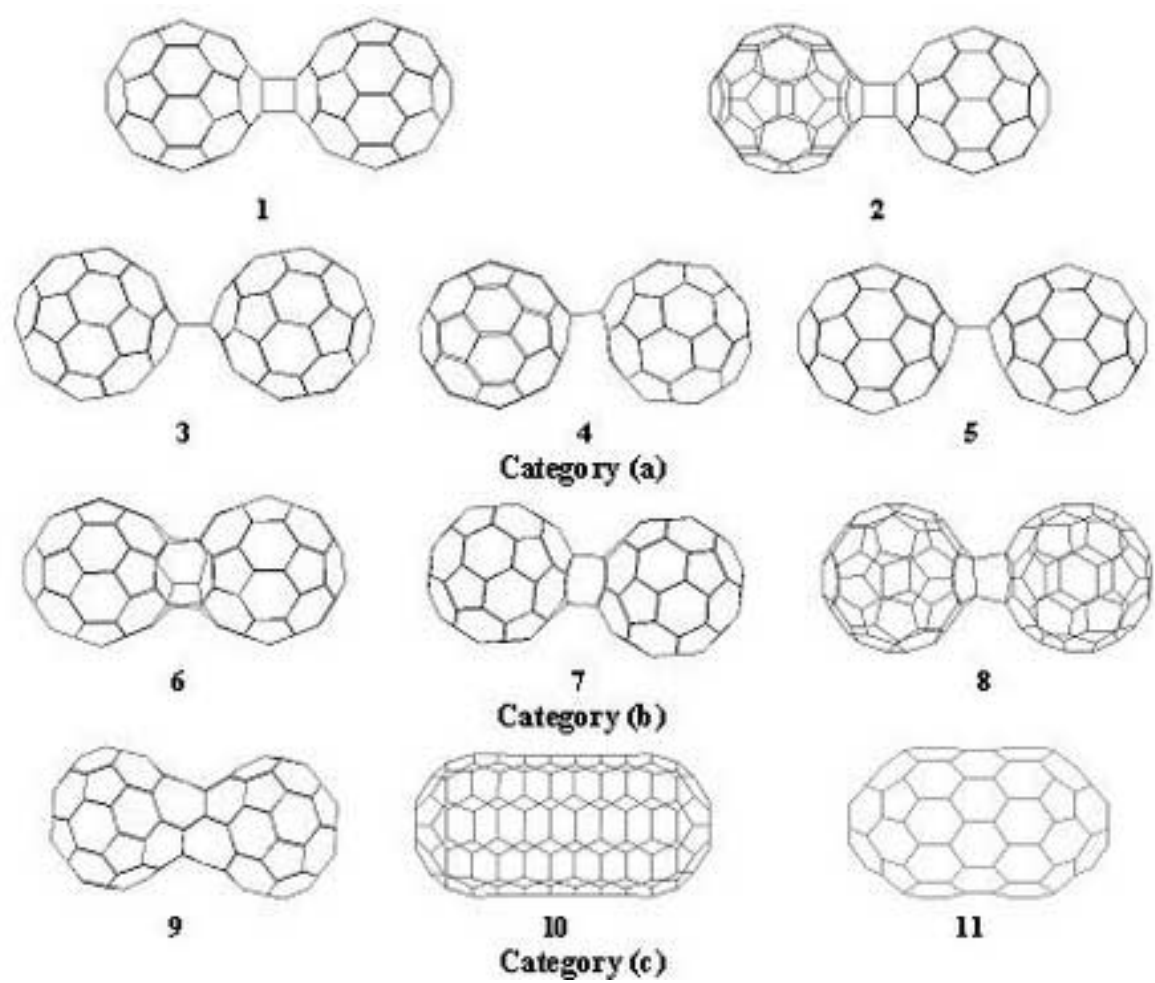

4
Category (a)

7
Category (b)

10
Category (c)

11